\documentclass[12pt,preprint]{aastex}

\slugcomment{Not to appear in Nonlearned J., 45.}

\shorttitle{}
\shortauthors{Wang et al.}

\begin{document}

\title{The role of active region coronal magnetic field in determining coronal mass ejection propagation direction}

\author{Rui Wang\altaffilmark{1}, Ying D. Liu\altaffilmark{1}, Xinghua Dai\altaffilmark{2,3}, Zhongwei Yang\altaffilmark{1}, Chong Huang\altaffilmark{4} and Huidong Hu\altaffilmark{1}}


\altaffiltext{1}{State Key Laboratory of Space Weather, National Space Science Center, Chinese Academy of Sciences, Beijing, China; liuxying@spaceweather.ac.cn}
\altaffiltext{2}{Key Laboratory of Solar Activity, National Astronomical Observatories, Chinese Academy of Sciences, Beijing 100012, China}
\altaffiltext{3}{University of Chinese Academy of Sciences, Beijing 100049, China}
\altaffiltext{4}{Institute of Space Sciences and School of Space Science and Physics, Shandong University, Weihai 264209, China}

\begin{abstract}
We study the role of the coronal magnetic field configuration of an active region in determining the propagation direction of a coronal mass ejection (CME). The CME occurred in the active region 11944 (S09W01) near the disk center on 2014 January 7 and was associated with an X1.2 flare. A new CME reconstruction procedure based on a polarimetric technique is adopted, which shows that the CME changed its propagation direction by around 28$^\circ$ in latitude within 2.5 R$_\odot$ and 43$^\circ$ in longitude within 6.5 R$_\odot$ with respect to the CME source region. This significant non-radial motion is consistent with the finding of M{\" o}stl et al. \citeyearpar{2015mostl}. We use nonlinear force-free field (NLFFF) and potential field source surface (PFSS) extrapolation methods to determine the configurations of the coronal magnetic field. We also calculate the magnetic energy density distributions at different heights based on the extrapolations. Our results show that the active region coronal magnetic field has a strong influence on the CME propagation direction. This is consistent with the ``channelling'' by the active region coronal magnetic field itself, rather than deflection by nearby structures. These results indicate that the active region coronal magnetic field configuration has to be taken into account in order to determine CME propagation direction correctly.

\end{abstract}





\section{INTRODUCTION}

Solar storms, known as coronal mass ejections (CMEs), are the most violent phenomenon in the solar atmosphere and a driver of major space weather effects. A reliable CME forecast has important consequences for life and technology on the Earth and in space. Accurate predictions of CME propagation direction are a major requirement for the forecast. However, CMEs do not always propagate in a constant direction and sometimes their propagation direction can be non-radial. CME propagation direction can deviate from the radial direction of the CME source region. Magnetic forces usually play an important role in deflecting CMEs within several solar radii. MacQueen et al. \citeyearpar{1986macQueen} indicated the influence of the background coronal magnetic field on CME deflection. Shen et al. \citeyearpar{2011shen} and Gui et al. \citeyearpar{2011gui} used gradients in the magnetic energy density of the solar corona to explain the observed deflection. Kay et al. \citeyearpar{2013kay,2015kay} developed a model called ForeCAT, which considers the influence of magnetic forces on CME propagation. Panasenco et al. \citeyearpar{2013panasenco} demonstrated that asymmetric large scale magnetic forces acting on CMEs can make CMEs propagate in the direction of the least resistance. The deflections can occur in both latitude and longitude \citep{2010liua,2010liub,2010lugaz,2013Isavnin}. Isavnin et al. \citeyearpar{2013Isavnin} reconstructed 15 CMEs between the minimum of Solar Cycle 23 and the maximum of Solar Cycle 24. They found that latitudinal deflections up 35$^\circ$ are observed while the maximum longitudinal deflection is only 5$^\circ$.4. Without out-of-ecliptic observations, more latitudinal deflection events were reported than longitudinal deflections \citep{1986macQueen,2009kilpua,2010Byrne,2013panasenco}. Byrne et al. \citeyearpar{2010Byrne} show a latitudinal deflection of $\sim$30$^\circ$ below 7 R$_\odot$ for the 2008 December 12 CME. With out-of-ecliptic observations provided by the Solar Orbiter satellite that will be launched into space in the near future, longitudinal deflection events will gain more attention.

According to previous studies, coronal holes where the fast solar wind flow comes from are thought to be one of the important factors that can deflect CMEs \citep{2006Cremades}. The coronal hole acts as a magnetic wall that constrains CME propagation. Kilpua et al. \citeyearpar{2009kilpua} suggested that CMEs cannot penetrate the magnetic wall and are guided by polar coronal hole fields to the equator. Gopalswamy et al. \citeyearpar{2009gopalswamy} considered the influence of the magnetic field strength of coronal holes and defined a coronal hole influence parameter (CHIP), whose direction has a good agreement with the direction of CME deflection. On the other hand, the CME moves away from nearby coronal holes and tends to deflect towards and propagate along a nearby helmet streamer or streamer belt (SB), for there are X-type magnetic structures (X-lines) where the magnetic stresses are minimum \citep{2009xie,2012zuccarello,2013kay}. Xie et al. (2009) found that slow CMEs (speed less than 400 km s$^{-1}$) tend to propagate towards and then along the SB due to the deflections by strong polar magnetic fields of corona holes. Kay et al. \citeyearpar{2013kay} suggested that magnetic background that contains strong magnetic gradients deflects CMEs to reach the SB; however for a weaker background, it just moves CMEs towards the SB but may not be capable of fully deflecting the CME along the SB. Panasenco et al. \citeyearpar{2013panasenco} and Kay et al. \citeyearpar{2013kay} both think about the CME deflection away from coronal holes and towards a streamer. Panasenco et al. \citeyearpar{2013panasenco} also looked at the rolling motion of filaments, which is similar to the deflection of CMEs.

As mentioned above, the driving forces originate from the magnetic field of the coronal holes outside the active region (AR), i.e., the CME source region. However, the role of the magnetic forces coming from the AR itself is not well understood. AR NOAA 11944 (S09W01) produced a fast halo CME with an X1.2 flare (S12W08) on 2014 January 7. This is an important event from a space weather perspective as forecasters expected a significant impact (a G3 class geomagnetic storm or higher) on the Earth, yet it did not occur. According to the study of M{\"o}stl et al. \citeyearpar{2015mostl}, the AR was near the disk center, but the CME propagated towards Mars, which was 51$^\circ$ away from the Earth in longitude. The deviation angle of the CME non-radial propagation in longitude was much larger than we thought. Most cases that have been reported have coronal longitudinal deflections $<\sim$ 20$^\circ$ \citep{2011gui,2013kay,2014Isavnin,2015kay}, although, there may be counterexamples in observations and simulations.
M{\" o}stl et al. \citeyearpar{2015mostl} showed that the distance of the coronal holes to the AR is large, which gives a CHIP value (defined by \citealt{2012Mohamed}) smaller than the one necessary to deflect a CME almost completely away from the Earth. They used a Graduate Cylindrical Shell (GCS) model \citep{2009thernisien} and a triangulation technique \citep{2010liua,2010liub} based on the STEREO data to estimate the CME propagation direction. Their results reveal that the change in CME propagation direction occurred close to the Sun and was likely to be related to the AR magnetic field rather than the coronal hole. They presented some global evidence to prove that the coronal magnetic field ``channelled'' the CME rather than deflected the CME. We follow their definition of the coronal ``channelling'', which is non-radial starting already with its inception on the Sun due to the AR magnetic fields. The ``deflection'' implies a change in direction, caused by the nearby structures, e.g., coronal holes and streamers. Here we present local evidence around the AR related to the coronal ``channelling'' as well as new CME reconstruction results. The topology of the magnetic field at different altitudes, especially near the CME source region, and the influence of magnetic stresses of the strong AR magnetic fields are both worth further studying.

In this paper, we use a polarimetric technique to reconstruct the CME and compare the results with those of M{\" o}stl et al. \citeyearpar{2015mostl}. We also investigate the configuration of coronal magnetic fields near the source region with a nonlinear force-free field (NLFFF) method and the magnetic field topology on a larger scale using the potential field source surface (PFSS) technique. Meanwhile, the changes of magnetic field energy density distribution with heights will be examined, which can demonstrate how magnetic forces play a significant role in channelling the CME.

\section{OBSERVATIONS AND ANALYSIS}\label{observations}
AR 11944 (S09W01) produced a very fast halo CME (linear speed of $\sim$1830 km s$^{-1}$) and an X1.2 flare (S12W08) on 2014 January 7. The flare started at 18:04 UT, peaked at 18:32 UT and ended at 18:58 UT. There exists another AR 11943 (S11W19) southwest of AR 11944. The CME source region is located around the border of the two ARs and belongs to AR 11944. At the time of the event, the two STEREO spacecraft were 151$^\circ$ ahead (A) and 153$^\circ$ behind (B) in heliospheric longitude with respect to the Earth and at distances of 0.96 AU (A) and 1.08 AU (B) from the Sun, respectively. Figure 1 shows a sided view of the CME from COR1 of STEREO-A and a backsided view from COR1 of STEREO-B. Note that the AR 11944 is located near the disk center as observed by the Atmospheric Imaging Assembly (AIA;~\citealt{2012lemen}) on board the Solar Dynamics Observatory (SDO;~\citealt{2012pesnell}). We can see that the CME was propagating to the southeast direction in STEREO-B. The non-radial propagation of the CME in longitude is more clearly shown in COR2 of STEREO-B where the CME was propagating to further distance. From the viewpoint of the Earth, the CME would be propagating to the southwest. M{\"o}stl et al. \citeyearpar{2015mostl} demonstrated that the CME propagated towards Mars, which was at a heliospheric longitude of 51$^\circ$ west of the Earth. They showed that the CME was deflected by 37$\pm$10$^\circ$ west of the source region or 45$\pm$10$^\circ$ away from the Sun-Earth line in heliospheric longitude \citep{2015mostl}. They further pointed out that the final direction of the CME was attained very close to the Sun, around 2.0 R$_\odot$ from the center of the Sun.

Figure 2a shows the EUV structures of the CME source region before the eruption occurred. Helioseismic and Magnetic Imager (HMI; \citealt{2012schou}) on board the SDO provides us high time and spatial resolution magnetograms with 0$^{\prime\prime}$.5 per pixel, as shown by the contour map overlaid on the EUV images. The positive polarity sunspot in the left side of Figure 2a has a very strong photospheric magnetic field, whose vertical component exceeds~3 kG. Livingston et al. \citeyearpar{2006livingston} found that the percentage of 12,804 sunspot groups between 1917 and 1974 with magnetic fields $\geq$3 kG is only 4.6\%. Some open EUV structures connected to the positive polarity sunspot are outlined by the red dotted lines and two lower EUV loops are outlined by the blue dotted lines. Figure 2b-2d show successive running difference images during the CME eruption, which better exhibit the evolution of the features in the kernel of the CME source region. The global solar disk observations of the erupting CME have been examined by M{\"o}stl et al. \citeyearpar{2015mostl}. The flare, coronal dimmings, post-eruption arcades, and a global coronal wave are thought to be the classical disk signatures of an erupting CME. The location of the front of the global coronal wave that is thought to be driven by the lateral expansion of the CME is visualized in their work. The global coronal wave at the onset phase can also be detected through the evolution of the features in Figure 2b-2d. The lower EUV loops in Figure 2a were blown away by the coronal wave propagating towards southwest. Between the strongest magnetic fields (red contours) at AR 11944 and the negative photospheric magnetic fields (cyan contours) at AR 11943, a possible magnetic flux rope (MFR) rose up then erupted with the wave, and was followed by a flare (see Figure 2d). In the west of the positive photospheric magnetic field (white), the coronal wave front is outlined by the red dashed line (see Figure 2c). The disk wave front likely corresponds to the CME front. According to the propagating direction of the wave, we can estimate that the CME was propagating away from the positive polarity sunspot with strong magnetic fields at the initial stage of the eruption. This is consistent with the coronagraph image in Figure 1b. Based on the evolution of the features above, we can roughly confirm the location of the CME source region as the same position of the flare.

\subsection{POLARIMETRIC RECONSTRUCTION OF THE CME}
Here we use a polarimetric technique to reconstruct the CME. This technique was developed by Moran et al. \citeyearpar{2004Moran} to reconstruct the 3D density distribution from the combination of intensity and polarized brightness images of LASCO. Based on the Thomson scattering theory, the polarization ratio determines the line of sight (LOS) averaged distances from the plane of sky (POS), which corresponds to the polarized image of CMEs as a reference surface for 3D polarimetric reconstruction. According to the reconstructed 3D location, each reconstructed CME pixel in the view of the equatorial plane can be located. The mass at each pixel depends on the position of each reconstructed pixel and the total brightness of the corresponding pixel. The mass at each pixel can be calculated using Equation (9) of Dai et al. \citeyearpar{2015dai}. The technique has been further improved for mass calculation of CMEs by Dai et al. \citeyearpar{2014dai,2015dai}. The advantage of the polarimetric reconstruction method is that, by using polarized observations from only a single viewpoint, it can reconstruct the mass distributions of CMEs. However, there exists ambiguities in the polarimetric analysis of CME observations \citep{2014dai}. Not all data can be utilized for reconstruction. For example, coronagraph images from STEREO-A show an implicit ambiguity \citep{2014dai}, and thus cannot be used. Figure 3a shows that the CME is split into two parts by the POS of STEREO-A (red dashed line). In the polarimetric technique, we cannot distinguish the sign of $\alpha$ in Equation (15) of Dai et al. \citeyearpar{2014dai}, which is the angle between the POS and a radial line connecting the center of the Sun and the average 3D location of electrons along the LOS of the CME pixel. Because of cos(-$\alpha$)=cos($\alpha$), $\alpha$ $\in$ [0$^\circ$, 90$^\circ$), the sign of $\alpha$ is ambiguous. The ambiguous locations of the electrons along the LOS of the CME pixel are denoted as the front and back of the POS corresponding to $\alpha$ and -$\alpha$. Thus we have to assume that the entire CME is either located before or behind the red dashed line. Therefore, although COR1 gives a sided view of the CME as shown in Figure 1a, it cannot be used for the polarimetric reconstruction due to this ambiguity. Fortunately, the viewpoint from STEREO-B has no such ambiguity because the entire CME is behind the POS of STEREO-B (see Figure 1), and thus it is reliable and appropriate for reconstruction.

The mass centers of the CME reconstructed from COR1 of STEREO-B at different times are also shown in Figure 3a. The green solid line represents the Sun-Earth line. The red and blue lines represent the LOS of STEREO-A and B, respectively. The biggest solid black sphere represents the mass center of the CME at the earliest time, i.e., 18:25 UT. The yellow lines are the radial lines connecting the mass centers with the center of the Sun. Table 1 gives the results of our reconstruction using data from COR1 of STEREO-B. Within 2.5 R$_\odot$, COR1 of STEREO-B provides us successive polarized data every 5 minutes. We use COR1 data because it is more accurate for its smaller field of view (FOV) (as shown in Figure 1) than COR2. After 18:45 UT, the CME exceeded the FOV of COR1 of STEREO-B, so we have to consider COR2 data. However, COR2 of STEREO-B cannot provide us polarized data until 20:00 UT (as shown in Figure 1c). After 20:00 UT, the CME exceeded the FOV of COR2. Therefore, for COR2 of STEREO-B only the polarized data around 20:00 UT can be used.
\begin{table}[t]\footnotesize
\epsscale{0.5}
\caption{CME reconstruction results at different times. The longitude is counted from the Sun-Earth line.}
\label{table:1}
\begin{tabular}{@{}lccccc@{}}
\tableline
\tableline
 Mass center         &Time (UT)   &Mass of CME (g)    &Heliocentric      &Longitude ($^\circ$) & Latitude ($^\circ$)   \\
                      &\verb|\|Instrument                      &                   &   distance (R$_\odot$)     &        &  \\
\tableline
 P1      &18:25\verb|\|B-COR1        &1.08e+15           &2.07         &22     &-37 \\
 P2      &18:30\verb|\|B-COR1        &2.68e+15           &2.24         &22     &-39 \\
 P3      &18:35\verb|\|B-COR1        &4.06e+15           &2.34         &25     &-40 \\
 P4      &18:40\verb|\|B-COR1        &5.11e+15           &2.29         &28     &-39 \\
 P5      &18:45\verb|\|B-COR1        &5.87e+15           &2.33         &29     &-38 \\
 P6      &20:00\verb|\|B-COR2        &1.10e+16           &6.54         &51     &-40 \\
\tableline
\end{tabular}
\end{table}
From 18:25 UT to 18:45 UT, the mass centers of the CME are at the heights between 2.0-2.5 R$_\odot$ relative to the center of the Sun (see Figures 3a). The mass of the CME keeps increasing with distance as expected. The heliospheric longitude of the CME mass center is slightly increasing but the latitude keeps around 40$^\circ$ (south). The CME deflection in heliospheric latitude can be seen from the coronagraph images of STEREO-B in Figure 1. Our results suggest that the CME was already deflected to a large longitude and latitude within 2.0 R$_\odot$. Compared with the approximate CME initial place (S12W08), the CME was deflected in heliospheric longitude by more than 20$^\circ$ below 2.5 R$_\odot$. Moreover, around 6.5 R$_\odot$ the longitudinal deflection reached around 40$^\circ$ with respect to the source region (also see Figures 3a and 3b). As shown by previous studies, most coronal longitudinal deflection events generally have a deflection $<\sim$20$^\circ$. The longitudinal deflection in the present case is unusually large in CME deflection observations.

Figure 3c and 3d show that the full CME white light structure in Figure 1c is well reconstructed by our polarimetric technique. Figure 3c shows the distribution of the reconstructed CME mass from COR2 of STEREO-B view. Figure 3d shows the reconstructed CME pixels in different heliocentric distance denoted by the color bar. The black solid polygon shows the boundary used for determining the CME mass center. When we determine the boundary above, it will introduce human error, but compared with the systematic uncertainty it is negligible. It is difficult to determine the systematic uncertainties inherent in our reconstruction method. However, our experience using the method suggests an error of about 10$^\circ$ for both the longitude and latitude. The results of M{\"o}stl et al. \citeyearpar{2015mostl} show that the CME is attained to W32$\pm$10$^\circ$ between 2.1 and 18.5 R$_\odot$ by means of the GCS method, and attained to W45$\pm$10$^\circ$ between 20 and 30 R$_\odot$ by means of the triangulation method. Our results are consistent with theirs below 2.5 R$_\odot$, but the CME has already been channelled into a direction around W50$^\circ$ at 6.54 R$_\odot$ (see Table 1). The reason is that the polarimetric method uses the mass center to show the position of a CME rather than the CME apex used by the GCS method. From Figure 3c we can see that a fraction of the mass is concentrated at the northeast of the Sun, which may result in the deviation from the results of M{\"o}stl et al. \citeyearpar{2015mostl}. For the reconstruction from COR1, it does not encounter this problem. On the other hand, Table 1 shows that the CME maintained around S40$^\circ$ in latitude beyond $\sim$2.0 R$_\odot$ but kept moving westward in longitude. This may be related to the large scale configurations of the magnetic field. The streamer belt looks like a ``U-shape" structure surrounding the source region and opens to the west. The CME moved to the south until reaching the top of the streamer belt (the X-line), then moved along the X-line. Therefore, the latitude kept constant beyond $\sim$2.0 R$_\odot$.

Our reconstruction results are consistent with those obtained by M{\"o}stl et al. \citeyearpar{2015mostl}, who used a forward modeling technique and a triangulation method. According to the synthesis of heliospheric observations of M{\"o}stl et al. \citeyearpar{2015mostl}, the CME arrived at Mars which was at a heliospheric longitude of 51$^\circ$ west of the Earth. From our CME reconstruction, we also see that the CME has already been channelled to a large angle at a relative low altitude. We think that the strong coronal magnetic field from the positive polarity sunspot is likely to play a role in determining the CME propagation direction. In the following section, we will investigate the configurations of the coronal magnetic field using two extrapolation techniques.

\subsection{CORONAL MAGNETIC FIELD RECONSTRUCTION}
We use different extrapolation models to reconstruct the coronal magnetic field within and surrounding the ARs. The simplest way to model the coronal magnetic field is to assume that the field is potential, i.e., no electric current. An AR, however, usually contains a large amount of electrical currents. The nonlinear force-free field (NLFFF) method is more appropriate in determining the coronal magnetic field configuration for an AR, while the PFSS technique may be more appropriate for a global field reconstruction.

Within $\sim$0.16 R$_\odot$ from the photosphere, we adopt the NLFFF method as proposed by \citet{2000wheatland} and extended by \citet{2004wiegelmann} and \citet{2010wiegelmann}. We use the NLFFF optimization code, which has been optimized for the application to SDO/HMI vector magnetograms \citep{2012wiegelmann}, to extrapolate the coronal magnetic field from the observed vector magnetograms in a Cartesian domain. A preprocessing procedure \citep{2006Wiegelmann} is employed to remove most of the net force and torque from the data, so the boundary can be more consistent with the force-free assumption. This extrapolation technique has been successfully applied in our previous work on two successive solar eruptions \citep{2014wang}. Here, we apply this NLFFF method to an area of 700$^{\prime\prime}$$\times$360$^{\prime\prime}$. The spatial resolution is 1$^{\prime\prime}$.0 per pixel. We adopt the Lambert Cylindrical Equal Area projected and remapped vector magnetic field data \citep{1990gary,2002calabretta,2006thompson} from HMI.

Figure 4 shows the NLFFF extrapolation results covering the main source region from different viewpoints. In order to distinguish different magnetic flux bundles, we use different colors to represent them (see Figure 4a). Compared with the EUV images of Figure 2b, the open magnetic flux bundles in black are consistent with the open EUV structures outlined in Figure 2b. The flux bundles in white are the core structure, which is shown as a fan-like structure inclining to the west. The flux bundles in blue are consistent with the post flare arch structures. The source region location is shown as the yellow star. As can be seen, the source region is located between the positive polarity sunspot and the open fan lines. The CME was probably affected by the strong magnetic pressure from the positive polarity sunspot and inclined to propagate along the open fan structures. However, we do not get the possible flux rope structure appearing in Figure 2a. A possible explanation is that this structure is very dynamic and therefore cannot be reconstructed with a model based on a static snapshot. In fact, we find some twisted field lines entangled with the lines originating from the sunspot with strong magnetic fields, which may evolve into this structure eventually.

From 1.16 to 2.5 R$_\odot$, the potential field source surface (PFSS) method \citep{1969schatten,2003Schrijver} is applied to reconstruct the coronal magnetic topology in a larger scale. Figure 5a gives the large scale structures of the magnetic field surrounding the ARs. The open magnetic field (cyan) is surrounded by the SB (purple). Figure 5b shows the open magnetic structures and some closed magnetic structures as part of the SB below 2.5 R$_\odot$. The open magnetic structures are similar to the NLFFF results at higher altitudes. One end of the open magnetic field lines roots in the strong positive magnetic field region, and the other end roots near the fan structure. The closed field lines are above the ARs as part of the SB and connect the positive polarity sunspot with strong photospheric magnetic fields to the negative polarity sunspot in the east of the AR 11944 (see Figure 4a). From Figure 5b we can see that the open magnetic structures are not along the radial direction but incline to the west. Figure 5c shows that the open field lines stretch out towards the north and south poles, respectively. The bundles of the open field lines stretching out to the south just pass through our reconstructed CME mass centers, which implies that the trajectory of the CME was likely along the open magnetic fields originating from the ARs.

The Lorentz force comprise a pressure force and a tension force. The tension force is the means to trap coronal plasma against the expansion of CMEs. As Figure 4b and c have shown, the closed magnetic field lines containing the tension force cover the source region and prevent the CME from expanding to higher places. The region of the open field lines, however, creates much less resistance, so the CME is more likely to pass through there.

Previous work \citep{2009xie,2012zuccarello,2013kay} indicated that the CME moves away from nearby coronal holes and tends to deflect towards and propagate along a nearby SB. For the distribution of the SB of this event, the CME propagated towards the SB in latitude but moved away from it in longitude (as shown in Figure 5a). According to Xie et al. \citeyearpar{2009xie}, there is no obvious correlation between the deflection of fast CMEs (V $\geq$ 400 km s$^{-1}$) and the SB. The remarkable difference of this CME deflection event from others is that there exists a sunspot with unusually strong magnetic fields nearby the erupting CME. This would produce a strong magnetic barrier and force the CME to propagate away from it. In order to evaluate the influence of the strong magnetic field on the CME propagation, we investigate the magnetic energy density distribution with altitudes in the following section.

\subsection{MAGNETIC FIELD ENERGY DENSITY DISTRIBUTION}
Previous studies showed that a nonuniform distribution of the background magnetic field energy density may cause the deflections of CMEs at an early stage and make CMEs tend to propagate to the region with lower magnetic energy density. Panasenco et al. \citeyearpar{2013panasenco} used magnetic stresses to represent this. The model of Kay et al. \citeyearpar{2013kay} for forecasting a CME's altered trajectory adopted magnetic pressure gradient. For the sake of simplicity, we present the magnetic energy density at some typical heights to study the role of the AR magnetic field in forcing the CME to propagate in a non-radial direction. The magnetic energy density $D_{me}$ is expressed as follows,
\begin{equation}
D_{me}=\frac{B^2}{2\mu_0}, \qquad \mu_0=4\pi\times10^{-7}W/A\centerdot m^{-1}
\end{equation}
where $B$ is in spherical coordinates, i.e., $B^2={B_\theta}^2+{B_\phi}^2+{B_r}^2$.

Figure 6 shows the magnetic energy density at 1.06 and 1.16 R$_\odot$, respectively. At such heights, the magnetic energy density comes from the results of the NLFFF extrapolation. The CME source region (S12W08) is located at the southwest of the positive polarity sunspot with strong magnetic fields (as shown in Figure 2). The distribution of the magnetic energy density shows a very strong gradient towards the southwest. This is produced by the presence of the strong sunspot. With the height increasing, the two-peak energy density distribution above the AR 11944 in Figure 6a gets into one peak in Figure 6b. The projections of the footpoints of the magnetic field lines traced from the CME mass centers (whose colors correspond to those in Figure 5) are located at the AR 11943.

Figure 7 gives the magnetic energy density distributions at higher altitudes, based on the PFSS extrapolation. Combining Figures 6 and 7, we find that the relative positions between the peaks of the density distributions (white contours) and the source region (the black cross) changed with the height. The magnetic field from the strong field region acts as a magnetic wall that constrains the CME propagation. Both Figures 6 and 7a show that the gradient descent direction points to the southwest, which is consistent with the CME propagation direction. Figure 7b shows that there appears another density peak in the northeast at 2.0 R$_\odot$. The gradient descent direction of the energy density distribution near the CME mass centers gradually changed from the south to the west. This gradient is consistent with the change of the projections of our reconstructed CME mass centers (from the colored crosses to the white one). Therefore, we think that the change in the CME propagation direction beyond $\sim$2.5 R$_\odot$ was likely caused by the presence of second density peak, while the change in the CME propagation direction within $\sim$2.5 R$_\odot$ probably resulted from the positive polarity sunspot with strong magnetic fields. The location of these open fields corresponds to the region of the lowest magnetic energy. The direction from the first density peak to the density valley is consistent with the propagation direction of the CME (from the black cross to the colored crosses; see Figure 7), which implies that the CME propagates to the less resistance region. Sum up, the change of the relative positions between the CME source region and the projections of the reconstructed CME mass centers is consistent with the change of distributions of the magnetic energy density with heights. This consistency supports the idea that the non-uniform distribution of the coronal magnetic field around the ARs plays a key role in channelling the CME. Note that we do not take into account any change in the magnetic field topology due to the passage of the CME. However, this method may still identify the environment that has influenced the CMEs channelling.
 \section{SUMMARY AND CONCLUSIONS} \label{conclusion}
In this work, we concentrate on the role of the active region coronal magnetic field in channelling a CME. A new CME reconstruction procedure based on a polarimetric technique is adopted. Our polarimetric reconstruction results imply that the CME was deflected close to the Sun and attained to its final direction within several solar radii. Our results also show that the CME was deflected in both latitude and longitude. The CME is demonstrated to be channelled into a direction around 28$^\circ$ (in latitude) away from the source region within 2.5 R$_\odot$. In longitude, it was channelled into a direction around 20$^\circ$ with respect to the source region within 2.5 R$_\odot$. This longitudinal deviation become about 40$^\circ$ from 2.5 to 6.5 R$_\odot$.

We use the NLFFF and PFSS methods to extrapolate the coronal magnetic field around the active regions at different heights. The NLFFF extrapolation results show a fan-like open magnetic field structure in the west of the source region. We obtained similar open field structures in higher corona from the PFSS model. We think that these open field structures probably provided a favorable channel for the CME propagation. In order to evaluate the influence of the magnetic pressure on the CME propagation, we examine the magnetic energy density distributions at different heights based on the NLFFF and PFSS extrapolation results. We compare the corresponding positions of the magnetic energy density peaks of each layer and the CME, as well as the relations between the gradient descent direction of the energy distribution and the changes of the CME mass centers. These comparisons indicate that the positive polarity sunspot on the northeast of the source region, which shows photospheric magnetic fields larger than 3 kG, plays a significant role in determining the CME propagation. There are two main factors that may have caused the CME non-radial propagation: (1) the magnetic pressure from the strong positive polarity sunspot forced the CME to the southwest; and (2) the configurations of the open and closed magnetic fields around the source region made the CME propagate to the least-resistance direction.

To conclude, this study shows the important role of the active region coronal magnetic field itself in determining the CME propagation direction. The active region coronal magnetic fields cause a significant change of CME propagation direction in both latitude and longitude. While our results are consistent with the ``channelled'' CME motion defined by M{\" o}stl et al. \citeyearpar{2015mostl}, they are obtained by a new approach and give a more complete story that is necessary to understand how the active region coronal magnetic field has channelled the CME at low altitudes. These results will help determine the propagation direction of CMEs correctly, which is important for space weather forecasting.

\acknowledgments

The work was supported by the Specialized Research Fund for State Key Laboratories of China, NSFC under grant 41374173, and the Recruitment program of Global Experts of China.

\clearpage


\begin{figure}
\epsscale{.99}
\plotone{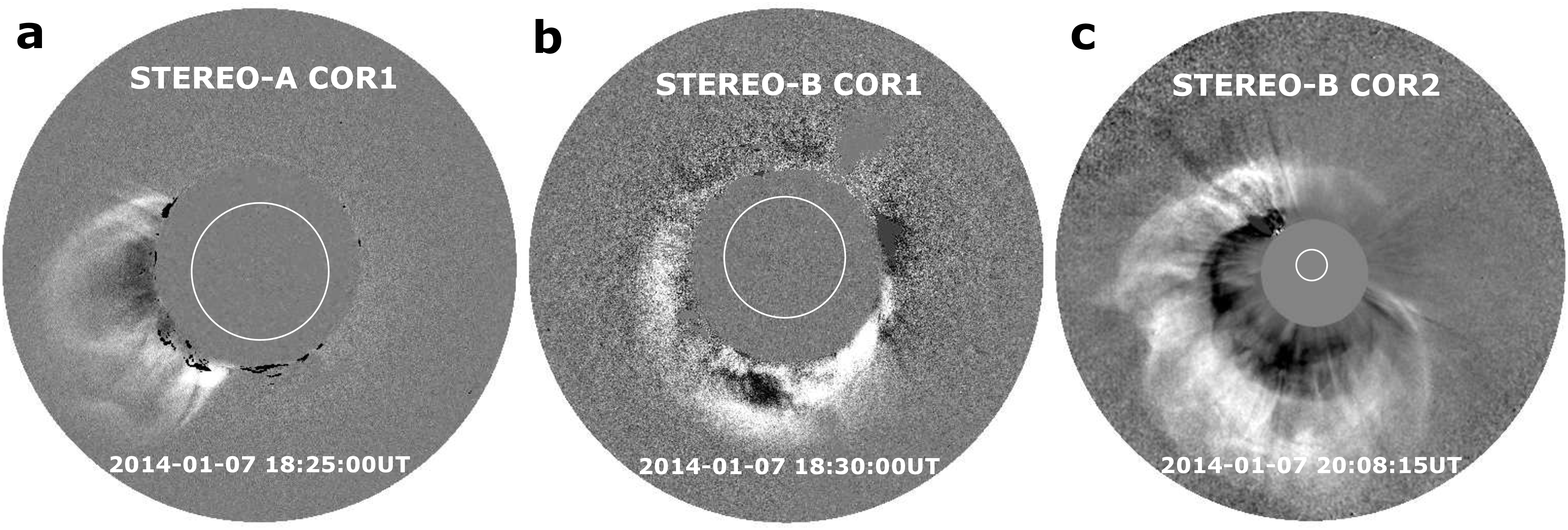}
\caption{Running difference images of the CME from STEREO-A and B. (a) A sided view of the CME from COR1 of STEREO-A. (b) A backsided view of the CME from COR1 of STEREO-B. (c) COR2 view of the CME from STEREO-B. The CME was propagating in the southeast direction for STEREO-B. The white circles mark the location and size of the solar disk. \label{figure1}}
\clearpage
\end{figure}
\begin{figure}
\epsscale{.99}
\plotone{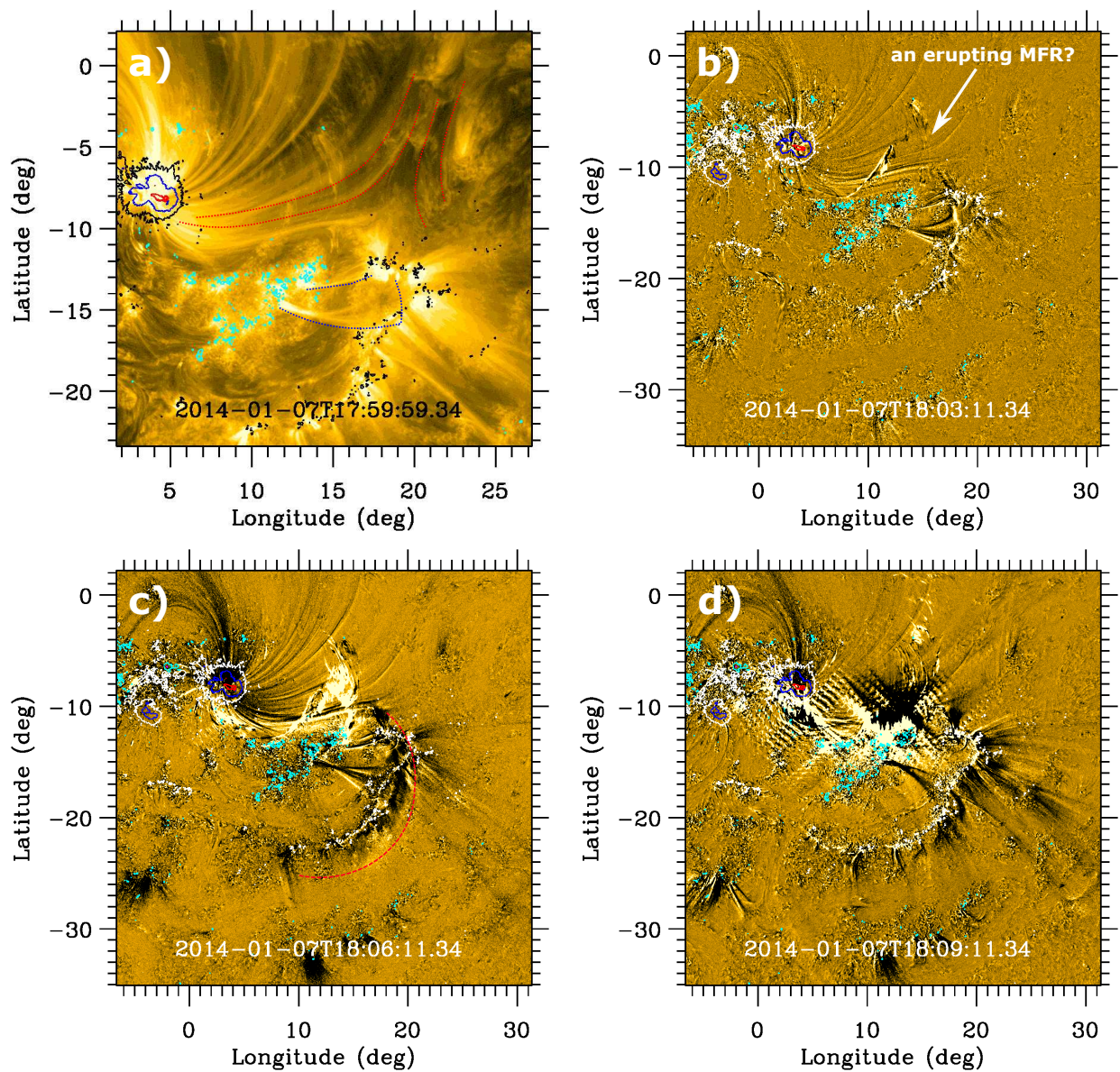}
\caption{The EUV structures and eruption of the CME source region from AIA 171 \AA. (a) EUV structures from AIA 171 \AA. The red and blue dotted lines mark open EUV structures and short EUV loops, respectively. (b)-(d) Successive running difference images of the eruption. The red dashed line marks the wave front of the CME. Contours of -1500, -500, 500, 1500, and 2500 G of the photospheric magnetic field along the line of sight from HMI magnetograms are overlaid on the EUV images and are marked in purple, cyan, black (a) (white (b)-(d)), blue and red, respectively. \label{figure2}}
\end{figure}
\begin{figure}
\epsscale{.84}
\plotone{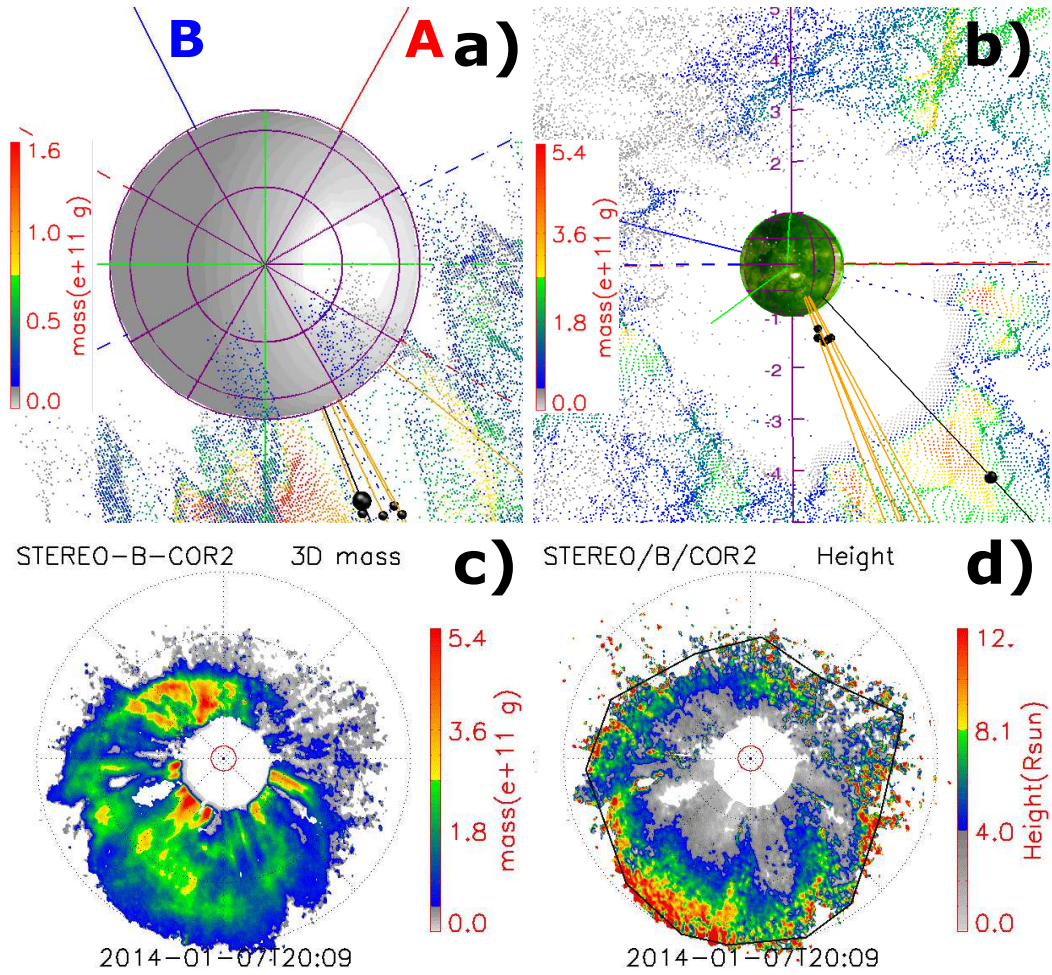}
\caption{Reconstruction of the 2014 January 7 CME from STEREO-B images. (a) Reconstructed CME mass distribution from COR1 of STEREO-B viewed from solar north. Green, red, and blue solid lines indicate the directions of the Earth, STEREO-A, and STEREO-B, respectively. The black solid and yellow lines show the directions of the mass centers (five black spheres) of the CME from 18:25 UT to 18:45 UT, respectively. The mass center at the earliest time is shown by the biggest black sphere. The dashed colored lines denote the corresponding plane of sky. The reconstruction results (color shading) at 18:25 UT are used. (b) Reconstructed CME mass distribution from COR2 of STEREO-B viewed from the Earth. The reconstruction results at 20:00 UT are used. The 171 \AA~image is mapped on the solar surface. The mass of each CME pixel is indicated by the color bar. (c) The distribution of the reconstructed CME mass from COR2 of STEREO-B view. (d) The distribution of the reconstructed CME pixels at different heliocentric distance (indicated by the color bar) from COR2 of STEREO-B view with error estimation as shown by the black solid polygon. The Sun is represented by the small red circle. \label{figure3}}
\end{figure}
\begin{figure}
\epsscale{.99}
\plotone{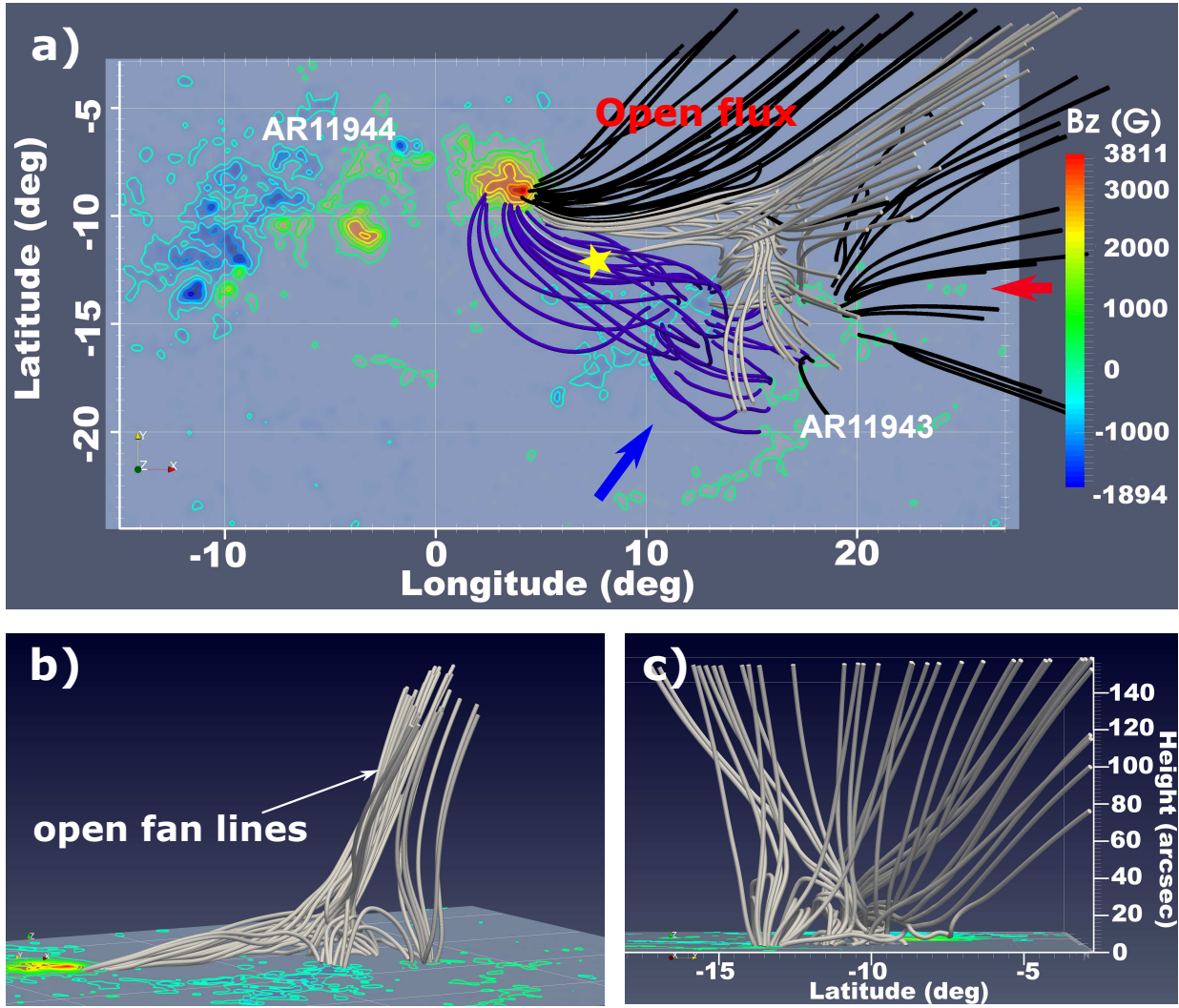}
\caption{NLFFF reconstruction of coronal magnetic fields of the source region at 18:00 UT. (a) Different magnetic flux bundles are represented by different color field lines. The yellow star marks the location of the CME source region. Color bar gives the vertical component of the background magnetic fields of the photosphere. (b) and (c) are viewed along the directions shown by the blue and red arrows in (a), respectively. \label{figure4}}
\end{figure}
\begin{figure}
\epsscale{.99}
\plotone{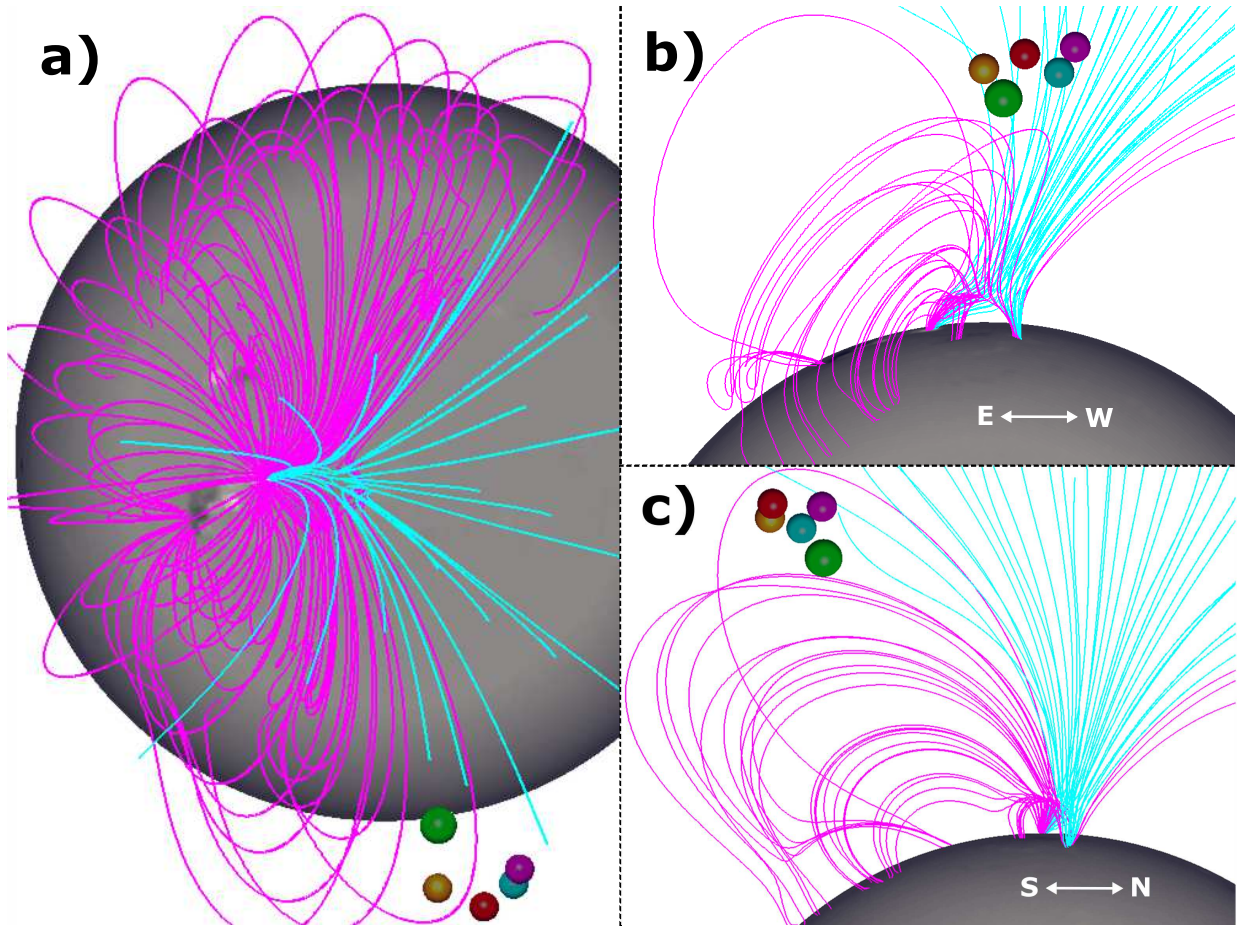}
\caption{PFSS reconstruction of the magnetic field above the source region. (a) Open magnetic fields (cyan) surrounded by the SB (purple) as viewed from the Earth. The observed magnetogram is mapped on the solar surface. The CME mass centers are represented by green, golden, red, cyan, and purple spheres, respectively. (b) Open magnetic field lines viewed from the south. Some closed magnetic field lines as part of the SB are also shown. (c) The same structures as shown in (b) but viewed from the west. \label{figure5}}
\end{figure}
\begin{figure}
\epsscale{.80}
\plotone{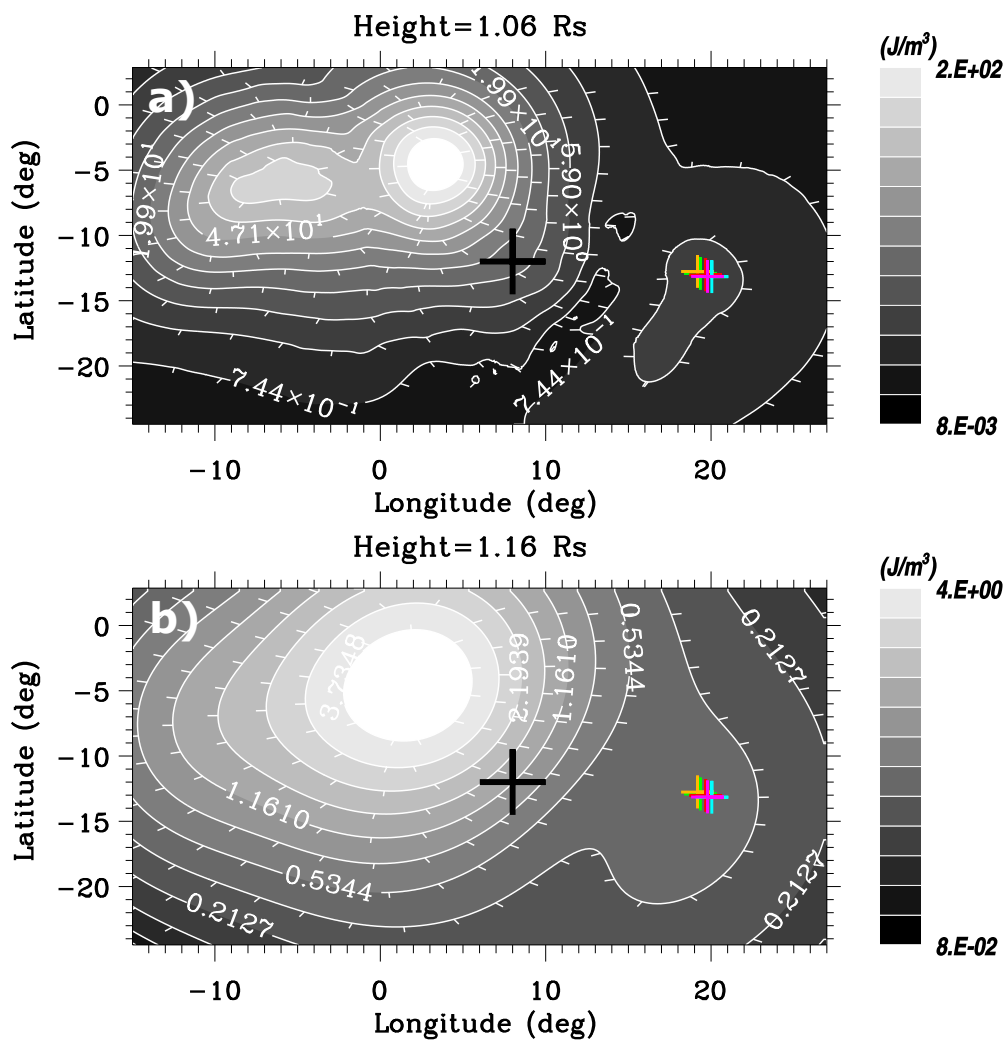}
\caption{Distribution of the magnetic energy density at the heights of 1.06 (upper panel) and 1.16 R$_\odot$ (lower panel) based on the NLFFF reconstruction. The black cross shows the projected location of the CME source region. The colored crosses (P1-P5) show the projections of the footpoints of the magnetic field lines traced from the reconstructed CME mass centers. \label{figure6}}
\end{figure}
\begin{figure}
\epsscale{.73}
\plotone{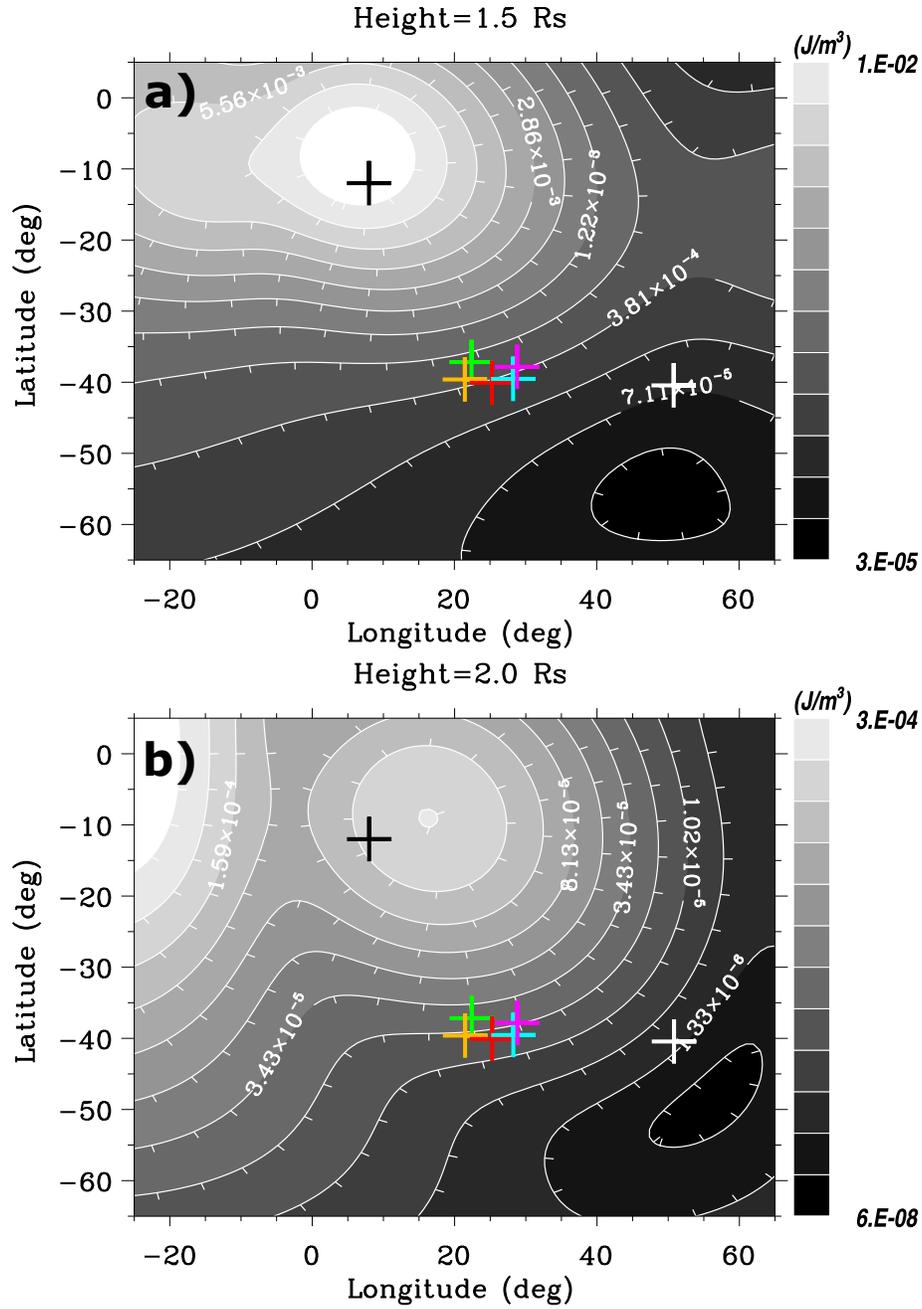}
\caption{Distribution of the magnetic energy density at the heights of 1.5 (upper panel) and 2.0 R$_\odot$ (lower panel) based on the PFSS reconstruction. The black cross shows the projected location of the source region. The colored (P1-P5) and white (P6) crosses show the projections of the reconstructed CME mass centers.  \label{figure7}}
\end{figure}

\clearpage




\end{document}